\documentclass[aip,jmp,preprint]{revtex4-1}

\begin{document}

\title{The Lorentz force equation as Fermi-Walker transport in geometrodynamics}
\author{Alcides Garat}
\affiliation{1. Instituto de F\'{\i}sica, Facultad de Ciencias,
Igu\'a 4225, esq. Mataojo, Montevideo, Uruguay.}

\date{August 15th, 2006}

\begin{abstract}
A new tetrad introduced within the framework of geometrodynamics
for non-null electromagnetic fields allows for the geometrical analysis of the Lorentz force equation and its solutions in curved spacetimes. When expressed in terms of this new tetrad, the electromagnetic field displays explicitly maximum simplification, and the degrees of freedom are manifestly revealed. In our manuscript we are deducing the Lorentz force equation on purely Riemannian geometrical grounds. The equation arises on the basis of Frenet-Serret analysis through the use of our new tetrads and gauge invariance arguments only. The force is deduced through a geometrical construction that precludes any other mathematical form other than the one already accepted. Therefore, a significant and fundamental result such as the first geometrical proof on the necessity of the force in the equation to have the structure already accepted in physics and not any other, is given. Through the use of the Frenet-Serret formulae and gauge invariance arguments we are also able to express in terms of the new tetrad vectors the Lorentz force equation as a generalized form of Fermi-Walker transport.
\end{abstract}

\pacs{}

\maketitle
\section{Introduction}
\label{introduction}

The new tetrads introduced in \cite{A}, yield maximum simplification in the expression of a non-null electromagnetic field in a curved spacetime. It is through this simplified expression of the electromagnetic field that we will get to the geometry associated with the Lorentz force equation and its solutions. It is the purpose of this manuscript three-fold. First, we are deducing the Lorentz force equation on purely Riemannian geometrical grounds. The equation arises on the basis of Frenet-Serret analysis through the use of our new tetrads and gauge invariance arguments only. The force is deduced through a geometrical construction that precludes any other mathematical form other than the one already accepted. Second, we will prove that the Lorentz force equation has its origin in Fermi-Walker transport, by making use of the electromagnetic tetrads already mentioned. Third, we will study different kinds of solutions to the Lorentz force equation from a geometric point of view through the use of the electromagnetic tetrads previously introduced. We use a metric with sign conventions $-+++$. If $F_{\mu\nu}$ is the electromagnetic field and $f_{\mu\nu}= (G^{1/2} / c^2) \: F_{\mu\nu}$ is the geometrized electromagnetic field, then the Einstein-Maxwell equations can be written,

\begin{eqnarray}
f^{\mu\nu}_{\:\:\:\:\:;\nu} &=& 0 \label{EM1}\\
\ast f^{\mu\nu}_{\:\:\:\:\:;\nu} &=& 0 \label{EM2}\\
R_{\mu\nu} &=& f_{\mu\lambda}\:\:f_{\nu}^{\:\:\:\lambda}
+ \ast f_{\mu\lambda}\:\ast f_{\nu}^{\:\:\:\lambda}\ , \label{EM3}
\end{eqnarray}

where  $\ast f_{\mu\nu}={1 \over 2}\:\epsilon_{\mu\nu\sigma\tau}\:f^{\sigma\tau}$
is the dual tensor of $f_{\mu\nu}$. The symbol $``;''$ stands for covariant derivative with respect to the metric tensor $g_{\mu\nu}$. At every point in spacetime there is a duality rotation by an angle $-\alpha$ that transforms a non-null electromagnetic field into an extremal field,

\begin{equation}
\xi_{\mu\nu} = e^{-\ast \alpha} f_{\mu\nu}\ = \cos(\alpha)\:f_{\mu\nu} - \sin(\alpha)\:\ast f_{\mu\nu}.\label{dref}
\end{equation}

The local scalar $\alpha$ is known as the complexion of the electromagnetic field. It is a local gauge invariant quantity. Extremal fields are essentially electric fields and they satisfy,

\begin{equation}
\xi_{\mu\nu} \ast \xi^{\mu\nu}= 0\ . \label{i0}
\end{equation}

They also satisfy the identity given by equation (64) in \cite{MW},

\begin{eqnarray}
\xi_{\alpha\mu}\:\ast \xi^{\mu\nu} &=& 0\ .\label{i1}
\end{eqnarray}

As antisymmetric fields, the extremal fields also verify the identity,

\begin{eqnarray}
\xi_{\mu\alpha}\:\xi^{\nu\alpha} -
\ast \xi_{\mu\alpha}\: \ast \xi^{\nu\alpha} &=& \frac{1}{2}
\: \delta_{\mu}^{\:\:\:\nu}\ Q \ ,\label{i2}
\end{eqnarray}

where $Q=\xi_{\mu\nu}\:\xi^{\mu\nu}=-\sqrt{T_{\mu\nu}T^{\mu\nu}}$
according to equations (39) in \cite{MW}. $Q$ is assumed not to be zero,
because we are dealing with non-null electromagnetic fields. $T_{\mu\nu} = \xi_{\mu\lambda}\:\:\xi_{\nu}^{\:\:\:\lambda}
+ \ast \xi_{\mu\lambda}\:\ast \xi_{\nu}^{\:\:\:\lambda}$ is the electromagnetic stress-energy tensor \cite{MW}.
In geometrodynamics, the Maxwell equations,

\begin{eqnarray}
f^{\mu\nu}_{\:\:\:\:\:;\nu} &=& 0 \label{L1}\nonumber\\
\ast f^{\mu\nu}_{\:\:\:\:\:;\nu} &=& 0 \ , \label{L2}
\end{eqnarray}

are telling us that two potential vector fields exist,

\begin{eqnarray}
f_{\mu\nu} &=& A_{\nu ;\mu} - A_{\mu ;\nu}\label{ER}\nonumber\\
\ast f_{\mu\nu} &=& \ast A_{\nu ;\mu} - \ast A_{\mu ;\nu} \ .\label{DER}
\end{eqnarray}

We can define then, a tetrad,

\begin{eqnarray}
U^{\alpha} &=& \xi^{\alpha\lambda}\:\xi_{\rho\lambda}\:A^{\rho} \:
/ \: (\: \sqrt{-Q/2} \: \sqrt{A_{\mu} \ \xi^{\mu\sigma} \
\xi_{\nu\sigma} \ A^{\nu}}\:) \label{UO}\\
V^{\alpha} &=& \xi^{\alpha\lambda}\:A_{\lambda} \:
/ \: (\:\sqrt{A_{\mu} \ \xi^{\mu\sigma} \
\xi_{\nu\sigma} \ A^{\nu}}\:) \label{VO}\\
Z^{\alpha} &=& \ast \xi^{\alpha\lambda} \: \ast A_{\lambda} \:
/ \: (\:\sqrt{\ast A_{\mu}  \ast \xi^{\mu\sigma}
\ast \xi_{\nu\sigma}  \ast A^{\nu}}\:)
\label{ZO}\\
W^{\alpha} &=& \ast \xi^{\alpha\lambda}\: \ast \xi_{\rho\lambda}
\:\ast A^{\rho} \: / \: (\:\sqrt{-Q/2} \: \sqrt{\ast A_{\mu}
\ast \xi^{\mu\sigma} \ast \xi_{\nu\sigma} \ast A^{\nu}}\:) \ .
\label{WO}
\end{eqnarray}

The four vectors (\ref{UO}-\ref{WO}) have the following algebraic properties,

\begin{equation}
-U^{\alpha}\:U_{\alpha}=V^{\alpha}\:V_{\alpha}
=Z^{\alpha}\:Z_{\alpha}=W^{\alpha}\:W_{\alpha}=1 \ .\label{TSPAUX}
\end{equation}

Using the equations (\ref{i1}-\ref{i2}) it is simple to prove that (\ref{UO}-\ref{WO}) are orthogonal. When we make the transformation,

\begin{eqnarray}
A_{\alpha} \rightarrow A_{\alpha} + \Lambda_{,\alpha}\ , \label{G1}
\end{eqnarray}

$f_{\mu\nu}$ remains invariant, and the transformation,

\begin{eqnarray}
\ast A_{\alpha} \rightarrow \ast A_{\alpha} +
\ast \Lambda_{,\alpha}\ , \label{G2}
\end{eqnarray}

leaves $\ast f_{\mu\nu}$ invariant,
as long as the functions $\Lambda$ and $\ast \Lambda$ are
scalars. Schouten defined what he called, a two-bladed structure
in a spacetime \cite{SCH}. These blades are the planes determined by the pairs
($U^{\alpha}, V^{\alpha}$) and ($Z^{\alpha}, W^{\alpha}$).
It was proved in \cite{A} that the transformation (\ref{G1}) generates a ``rotation'' of the tetrad vectors ($U^{\alpha}, V^{\alpha}$) into ($\tilde{U}^{\alpha}, \tilde{V}^{\alpha}$) such that these ``rotated'' vectors ($\tilde{U}^{\alpha}, \tilde{V}^{\alpha}$) remain in the plane or blade one generated by ($U^{\alpha}, V^{\alpha}$).
It was also proved in \cite{A} that the transformation (\ref{G2}) generates a ``rotation'' of the tetrad vectors ($Z^{\alpha}, W^{\alpha}$) into ($\tilde{Z}^{\alpha}, \tilde{W}^{\alpha}$) such that these ``rotated'' vectors ($\tilde{Z}^{\alpha}, \tilde{W}^{\alpha}$) remain in the plane or blade two generated by ($Z^{\alpha}, W^{\alpha}$).
Once we introduced all these elements found in \cite{A}, we state the purpose of this manuscript. A particle of charge $q$ and mass $m$ in the presence of the electromagnetic field $f_{\mu\nu}$ is going to follow a timelike curve determined by the Lorentz force equation \cite{ROH}$^{,}$\cite{JAW}$^{,}$\cite{DeWB}$^{,}$\cite{DeWCM}$^{,}$\cite{MC1}. Through gauge invariance arguments applied to Frenet-Serret formulae we are able to deduce the Lorentz force equation in a curved Lorentzian spacetime. It will be proved that the set of all curves that satisfy the Lorentz force equation can be generated through a generalized form of Fermi-Walker transport making use of the tetrad (\ref{UO}-\ref{WO}). This paper is organized as follows. In section (\ref{FermiWalker}) we introduce the concept of Fermi-Walker transport adapted to the problem of geometrodynamics. In section (\ref{Lorentzequation}) we deduce the Lorentz force equation through the use of the Frenet-Serret expressions and gauge invariance arguments using two different approaches. We also prove that it represents a generalized form of Fermi-Walker transport. In section (\ref{anotherview}) we analyze the possibility of having special kinds of solutions to the Lorentz force equation such as pure blade one solutions. Finally in section (\ref{anothergenview}) it is our objective to find special congruences of solutions to the Lorentz force equation, as a more general example. These are new geometrical techniques that we are introducing in order to analyze the Lorentz force equation as a differential equation in a curved spacetime.


\section{Fermi-Walker transport in geometrodynamics}
\label{FermiWalker}

The Frenet-Serret formulae for a tetrad associated for instance with a timelike curve $x_{t}^{\beta}(s)$ in spacetime \cite{SY}$^{,}$\cite{HS} are,

\begin{eqnarray}
{\delta A^{\alpha} \over \delta s} &=& b \: B^{\alpha} \label{A}\\
{\delta B^{\alpha} \over \delta s} &=& c \: C^{\alpha} + b \: A^{\alpha} \label{B}\\
{\delta C^{\alpha} \over \delta s} &=& d \: D^{\alpha} - c \: B^{\alpha} \label{V}\label{C}\\
{\delta D^{\alpha} \over \delta s} &=& -d \: C^{\alpha} \label{D} \ ,
\end{eqnarray}

where $ A^{\alpha}( x^{\beta}(s)) = {dx_{t}^{\alpha} \over ds}$ is the unit timelike tangent vector to the curve. The triad $( B^{\alpha}, C^{\alpha}, D^{\alpha})$ is formed by unit spacelike vector fields. They are known as the first, second and third normals. $b$, $c$ and $d$ are known as the first, second and third curvatures \cite{SY}. We know that when the local scalar functions $b$, $c$ and $d$ are equal to zero, the curve is a geodesic \cite{MM}. If $b \neq 0$, then we can define the Fermi-Walker transport of a vector $F^{\alpha}$ as,

\begin{equation}
{\delta F^{\alpha} \over \delta s} =  b \: F_{\rho} \: (A^{\alpha} \: B^{\rho} - A^{\rho} \: B^{\alpha}) \ , \label{FWT}\\
\end{equation}

where $B^{\alpha}$ and $b$ are the first normal and first curvature to the timelike curve $x^{\beta}_{t}(s)$. The ``absolute derivative'' ${\delta \over \delta s}$ is defined in \cite{SY}. For instance, along a timelike curve with unit tangent vector $A^{\alpha}$,

\begin{equation}
{\delta F^{\alpha} \over \delta s} =  F^{\alpha}_{\:\:\: ;\rho} \: A^{\rho} \ . \label{AD}\\
\end{equation}

Analogously, Fermi-Walker transport of a vector $F^{\alpha}$ can be defined along a spacelike curve as \cite{SY},

\begin{equation}
{\delta F^{\alpha} \over \delta s} =  - d \: F_{\rho} \: (C^{\alpha} \: D^{\rho} - C^{\rho} \: D^{\alpha}) \ . \label{FWS}\\
\end{equation}

Fermi-Walker transport given by (\ref{FWS}) is along the spacelike curve $x^{\beta}_{s}(s)$, whose unit spacelike tangent vector is $D^{\alpha}$. In \cite{A} we defined a new orthonormal tetrad, expressions (\ref{UO}-\ref{WO}) for spacetimes where non-null electromagnetic fields are present $ (U^{\alpha}, V^{\alpha}, Z^{\alpha}, W^{\alpha})$. Written in terms of these tetrad vectors, the electromagnetic field is,

\begin{equation}
f_{\alpha\beta} = -2\:\sqrt{-Q/2}\:\:\cos\alpha\:\:U_{[\alpha}\:V_{\beta]} +
2\:\sqrt{-Q/2}\:\:\sin\alpha\:\:Z_{[\alpha}\:W_{\beta]}\ .\label{EMF}
\end{equation}

It is precisely the tetrad structure of (\ref{EMF}) that allows for a ``connection'' to the Fermi-Walker transport. It is also the possibility of ``rotating'' through a scalar angle $\phi$ on blade one, the tetrad vectors $U^{\alpha}$ and $V^{\alpha}$, such that
$U_{[\alpha}\:V_{\beta]}$ remains invariant \cite{A}, that we have the freedom to build timelike curve congruences as a function of the ``angle'' $\phi$ on blade one. Similar on blade two. A ``rotation'' of the tetrad vectors $Z^{\alpha}$ and $W^{\alpha}$ through an ``angle'' $\varphi$, such that
$Z_{[\alpha}\:W_{\beta]}$ remains invariant \cite{A}, entails the freedom to build spacelike curve congruences as a function of the ``angle'' $\varphi$ on blade two. We must remember that to each local function $\phi$ there corresponds, through an isomorphic relationship, an electromagnetic gauge transformation local scalar $\Lambda$ as a counterpart to $\phi$ on blade one \cite{A}. Similar for $\varphi$ on blade two. That is, an electromagnetic local gauge transformation of the vectors $(U^{\alpha}, V^{\alpha})$ generates a ``rotation'' of these vectors on blade one at every point in spacetime, such that these vectors remain on blade one after the ``rotation''. We called the Abelian group of ``rotations'' on blade one, $LB1$. There are also discrete transformations on blade one, but we are going to call all of the LB1 transformations generically as ``rotations''. Similar for $(Z^{\alpha}, W^{\alpha})$ on blade two, under the group $LB2$.

Summarizing the above ideas, we are going to find for each tetrad, solution to the Einstein-Maxwell equations, congruences of timelike curves, solutions to the Lorentz force equation.

\section{The Lorentz force equation}
\label{Lorentzequation}

We are going to consider a timelike curve in spacetime, such that at each point, the tangent to this curve is given by the unit timelike vector $ u^{\alpha} = u\:U^{\alpha}_{(\phi)} + v\:V^{\alpha}_{(\phi)} + z\:Z^{\alpha}_{(\varphi)} + w\:W^{\alpha}_{(\varphi)}$; $\:\:\:u^{\alpha}\: u_{\alpha}  = -1$. The functions $u$, $v$, $z$ and $w$ are local scalars. The ``angles'' $(\phi,\varphi)$ are also local scalars and we are introducing them in order to express $u^{\alpha}$ in the most general way. The tetrad vectors $ (U^{\alpha}, V^{\alpha}, Z^{\alpha}, W^{\alpha})$ are provided through (\ref{UO}-\ref{WO}). Simultaneously, we are going to consider the ``rotated'' tetrad $(\tilde{U}^{\alpha}, \tilde{V}^{\alpha}, \tilde{Z}^{\alpha}, \tilde{W}^{\alpha})$ defined below through expressions (\ref{uUO}-\ref{uWO}). For the sake of simplicity we are going to assume that the transformation of the two vectors $(U^{\alpha},\:V^{\alpha})$ on blade one, given in (\ref{UO}-\ref{VO}) by the ``angle'' $\phi$ is a proper transformation, that is, a boost. For discrete improper transformations the result follows the same lines \cite{A}. Therefore we can write,

\begin{eqnarray}
U^{\alpha}_{(\phi)}  &=& \cosh(\phi)\: U^{\alpha} +  \sinh(\phi)\: V^{\alpha} \label{UT} \\
V^{\alpha}_{(\phi)} &=& \sinh(\phi)\: U^{\alpha} +  \cosh(\phi)\: V^{\alpha} \label{VT} \ .
\end{eqnarray}

The transformation of the two tetrad vectors $(Z^{\alpha},\:W^{\alpha})$ on blade two, given in (\ref{ZO}-\ref{WO}) by the ``angle'' $\varphi$, can be expressed as,

\begin{eqnarray}
Z^{\alpha}_{(\varphi)}  &=& \cos(\varphi)\: Z^{\alpha} -  \sin(\varphi)\: W^{\alpha} \label{ZT} \\
W^{\alpha}_{(\varphi)}  &=& \sin(\varphi)\: Z^{\alpha} +  \cos(\varphi)\: W^{\alpha} \label{WT} \ .
\end{eqnarray}

We can define then, a ``rotated'' orthonormal tetrad with respect to the tetrad (\ref{UO}-\ref{WO}), along the integral timelike curve as,

\begin{eqnarray}
\tilde{U}^{\alpha} &=& \xi^{\alpha\lambda}\:\xi_{\rho\lambda}\:u^{\rho} \:
/ \: (\: \sqrt{-Q/2} \: \sqrt{u_{\mu} \ \xi^{\mu\sigma} \
\xi_{\nu\sigma} \ u^{\nu}}\:) \label{uUO}\\
\tilde{V}^{\alpha} &=& \xi^{\alpha\lambda}\:u_{\lambda}  \:
/ \: (\:\sqrt{u_{\mu} \ \xi^{\mu\sigma} \
\xi_{\nu\sigma} \ u^{\nu}}\:) \label{uVO}\\
\tilde{Z}^{\alpha} &=& \ast \xi^{\alpha\lambda} \:  u_{\lambda} \:
/ \: (\:\sqrt{ u_{\mu}  \ast \xi^{\mu\sigma}
\ast \xi_{\nu\sigma}  u^{\nu}}\:)
\label{uZO}\\
\tilde{W}^{\alpha} &=& \ast \xi^{\alpha\lambda}\: \ast \xi_{\rho\lambda}
\: u^{\rho} \: / \: (\:\sqrt{-Q/2} \: \sqrt{ u_{\mu}
\ast \xi^{\mu\sigma} \ast \xi_{\nu\sigma}  u^{\nu}}\:) \ ,
\label{uWO}
\end{eqnarray}

where $u^{\alpha}$ is the tangent along the timelike curve. We must be careful about the definition of tetrad vectors (\ref{uZO}-\ref{uWO}). There is the possibility that the timelike tangent vector $u^{\alpha}$ does not have non-zero components on blade two. We solve this particular problem as an example in the next section. As a comment we can say that if we replace the two tetrad vectors (\ref{uZO}-\ref{uWO}) for the vectors (\ref{ZO}-\ref{WO}), we would have an orthonormal tetrad too. The tetrads in their construction involve a skeleton built with extremal fields, and two vector fields $X^{\mu}$ and $Y^{\mu}$ which are gauge vectors. The gauge vectors can be chosen freely if we are not producing a trivial vector, of course. Any choice corresponds to a gauge transformation (Lorentz transformation) of the tetrad vectors with gauge vector choice given by the usual potentials $X^{\mu}=A^{\mu}$, $Y^{\mu}=*A^{\mu}$, see reference \cite{A}. Any other choice of gauge vectors represents a gauge transformation with respect to the pair of tetrad vectors that define either blade one or two with gauge vectors given by the usual potentials $X^{\mu}=A^{\mu}$, $Y^{\mu}=*A^{\mu}$. Therefore, when we chose the gauge vectors to be the velocity $X^{\mu} = Y^{\mu} = u^{\mu}$, it represents a gauge choice. It happens that this particular gauge choice is absolutely convenient to prove the geometrical origin of the force in the Lorentz force equation. At the same time , the antisymmetrical combinations of the two generating tetrad vectors on either blade $U^{[\alpha}\:V^{\beta]}$ and $Z^{[\alpha}\:W^{\beta]}$ are gauge invariant (Lorentz invariant), that is invariant under transformations (\ref{G1}-\ref{G2}) or equivalently (\ref{UT}-\ref{WT}). They represent in tetrad language, the geometrical version of the standard gauge invariance of the electromagnetic field. Again, we are exploiting repeatedly this freedom in order to prove the Fermi-Walker nature of the Lorentz force equation.

\subsection{Frenet-Serret first version }
\label{FSfv}

We proceed then, to write the Frenet-Serret expressions associated to the timelike curve whose tangent is $u^{\alpha}$.

\begin{eqnarray}
{\delta u^{\alpha} \over \delta s}  &=& A\:\left(u_{\beta}\: \xi^{\alpha\beta}\right) + B\:\left(u_{\beta}\: \ast \xi^{\alpha\beta}\right) + C\: \left( (u_{\beta}\:\tilde{W}^{\beta})\:\tilde{U}^{\alpha} - (u_{\beta}\:\tilde{U}^{\beta})\:\tilde{W}^{\alpha}  \right)
\label{FSE} \ ,
\end{eqnarray}

where $ \tilde{U}^{\alpha}$ and $ \tilde{W}^{\beta}$ are provided by expressions (\ref{uUO}) and (\ref{uWO}). The functions $A$, $B$ and $C$ are local scalars. It is very simple to prove that $u_{\alpha}$ is orthogonal to $u_{\beta}\: \xi^{\alpha\beta}$, $u_{\beta}\: \ast \xi^{\alpha\beta}$, and $(u_{\beta}\:\tilde{W}^{\beta})\:\tilde{U}^{\alpha} - (u_{\beta}\:\tilde{U}^{\beta})\:\tilde{W}^{\alpha}$. It is also very simple to prove using the tetrad vectors (\ref{uUO}-\ref{uWO}) and the identities (\ref{i1}-\ref{i2}) that the three vectors $u_{\beta}\: \xi^{\alpha\beta}$, $u_{\beta}\: \ast \xi^{\alpha\beta}$, and $(u_{\beta}\:\tilde{W}^{\beta})\:\tilde{U}^{\alpha} - (u_{\beta}\:\tilde{U}^{\beta})\:\tilde{W}^{\alpha}$ are orthogonal among them.
The extremal field and its dual can be written as \cite{A},

\begin{eqnarray}
\xi_{\alpha\beta} &=& -2\:\sqrt{-Q/2}\:U_{[\alpha}\:V_{\beta]}\label{ET}\\
\ast \xi_{\alpha\beta} &=& 2\:\sqrt{-Q/2}\:Z_{[\alpha}\:W_{\beta]}\ .\label{DET}
\end{eqnarray}

Because of gauge invariance we also know that \cite{A},

\begin{eqnarray}
U^{[\alpha}_{(\phi)}\:V^{\beta]}_{(\phi)} = U^{[\alpha}\:V^{\beta]} = \tilde{U}^{[\alpha}\:\tilde{V}^{\beta]} \label{GEQ1}\\
Z^{[\alpha}_{(\varphi)}\:W^{\beta]}_{(\varphi)} = Z^{[\alpha}\:W^{\beta]} = \tilde{Z}^{[\alpha}\:\tilde{W}^{\beta]} \label{GEQ2} \ .
\end{eqnarray}

We would like to emphasize that expression $\left( \tilde{W}^{\beta}\:\tilde{U}^{\alpha} - \tilde{W}^{\alpha}\:\tilde{U}^{\beta} \right)$ is not invariant under gauge transformations of the tetrad vectors. $u^{\alpha}$ does not have a gauge dependence, $\xi_{\alpha\beta}$ and $\ast \xi_{\alpha\beta}$ are both gauge invariant, and so are $A$, $B$ and $C$. The Lorentz force equation does not depend on gauge. Nonetheless $\left( \tilde{W}^{\beta}\:\tilde{U}^{\alpha} - \tilde{W}^{\alpha}\:\tilde{U}^{\beta} \right)$ is gauge dependent, therefore the only possible gauge invariant choice for $C$, is $C=0$. Therefore, we have a unique justification for considering the curvature scalar $C = 0$. Then, following the line of thinking in \cite{SY}, we can consider or make the choice for the curvature associated gauge invariant scalars to be, $A = (q/m)\:\cos(\alpha)$, $B = (q/m)\:\sin(\alpha)$ and $C = 0$ where $q$ is the charge and $m$ is the mass of the particle following the integral timelike curve to which $u^{\alpha}$ is tangent.  Then, we can rewrite (\ref{FSE}) as,

\begin{eqnarray}
{\delta u^{\alpha} \over \delta s}  &=& -(q/m)\:\sqrt{-Q/2}\:\cos(\alpha) \:u_{\beta}\: \left(U^{\alpha}\:V^{\beta}- U^{\beta}\:V^{\alpha}\right) + \nonumber\\
&&(q/m)\:\sqrt{-Q/2}\:\sin(\alpha)\:u_{\beta}\: \left(Z^{\alpha}\:W^{\beta}- Z^{\beta}\:W^{\alpha}\right) \label{FSESIMPALPHA} \ .
\end{eqnarray}

Making use of the expression for the electromagnetic field (\ref{EMF}) we can write (\ref{FSESIMPALPHA}) as,

\begin{eqnarray}
{\delta u^{\alpha} \over \delta s}  = (q/m)\:u_{\beta}\:f^{\alpha\beta}  \label{FSEMF} \ .
\end{eqnarray}

It is clear that equation (\ref{FSEMF}) was deduced on the only assumption of gauge invariance for the equation describing the timelike trajectory of a particle through the use of a Frenet-Serret formulae. We must stress that we are using the term Frenet-Serret in a general way, since the vectors involved to the right of equation (\ref{FSE}) are not normalized.

\subsection{Frenet-Serret second version }
\label{FSsv}

We can engineer another way of writing the right hand side of equation (\ref{FSE}),

\begin{eqnarray}
{\delta u^{\alpha} \over \delta s}  &=& A\:u_{\beta}\: f^{\alpha\beta} + B\:\left( (u_{\beta}\:f^{\gamma\beta}\:\tilde{V}_{\gamma})\:\tilde{Z}^{\alpha} - (u_{\beta}\:f^{\gamma\beta}\:\tilde{Z}_{\gamma})\:\tilde{V}^{\alpha}\right) + \nonumber\\ &&C\: \left( (u_{\beta}\:\tilde{W}^{\beta})\:\tilde{U}^{\alpha} - (u_{\beta}\:\tilde{U}^{\beta})\:\tilde{W}^{\alpha} \right)
\label{FSE2} \ .
\end{eqnarray}

We can observe the following. $u^{\alpha}$ is orthogonal to $\:u_{\beta}\: f^{\alpha\beta}$, $\left((u_{\beta}\:f^{\gamma\beta}\:\tilde{V}_{\gamma})\:\tilde{Z}^{\alpha} - (u_{\beta}\:f^{\gamma\beta}\:\tilde{Z}_{\gamma})\:\tilde{V}^{\alpha}\right)$, and to $\left( (u_{\beta}\:\tilde{W}^{\beta})\:\tilde{U}^{\alpha} - (u_{\beta}\:\tilde{U}^{\beta})\:\tilde{W}^{\alpha} \right)$. Next we observe that $\:u_{\beta}\: f^{\alpha\beta}$ is orthogonal to \\ $\left( (u_{\beta}\:f^{\gamma\beta}\:\tilde{V}_{\gamma})\:\tilde{Z}^{\alpha} - (u_{\beta}\:f^{\gamma\beta}\:\tilde{Z}_{\gamma})\:\tilde{V}^{\alpha}\right)$ and $\left( (u_{\beta}\:\tilde{W}^{\beta})\:\tilde{U}^{\alpha} - (u_{\beta}\:\tilde{U}^{\beta})\:\tilde{W}^{\alpha} \right)$. Finally we observe that $\left( (u_{\beta}\:f^{\gamma\beta}\:\tilde{V}_{\gamma})\:\tilde{Z}^{\alpha} - (u_{\beta}\:f^{\gamma\beta}\:\tilde{Z}_{\gamma})\:\tilde{V}^{\alpha}\right)$ is orthogonal to $\left( (u_{\beta}\:\tilde{W}^{\beta})\:\tilde{U}^{\alpha} - (u_{\beta}\:\tilde{U}^{\beta})\:\tilde{W}^{\alpha} \right)$. All of these orthogonalities proved through the antisymmetry of several tensors, the iterative use of equations (\ref{i0}-\ref{i2}), expression (\ref{EMF}) for the electromagnetic field, expressions (\ref{ET}-\ref{DET}) for the extremal field and its dual, and the orthogonality of the normalized vectors (\ref{uUO}-\ref{uWO}). It is straightforward to prove that the objects $\left( (u_{\beta}\:f^{\gamma\beta}\:\tilde{V}_{\gamma})\:\tilde{Z}^{\alpha} - (u_{\beta}\:f^{\gamma\beta}\:\tilde{Z}_{\gamma})\:\tilde{V}^{\alpha}\right)$ and $\left( (u_{\beta}\:\tilde{W}^{\beta})\:\tilde{U}^{\alpha} - (u_{\beta}\:\tilde{U}^{\beta})\:\tilde{W}^{\alpha} \right)$ are not gauge invariant under transformations analogous to (\ref{UT}-\ref{WT}) for the tetrad vectors (\ref{uUO}-\ref{uWO}). But we know again that the Lorentz force equation is gauge invariant, and the only way for this property to hold is for the curvature scalars $B$ and $C$ to take the value $B=C=0$. This time following the line of thinking in \cite{SY} we can make the choice $A = q/m$. Again through gauge invariance arguments we found the Lorentz force equation. The only curvature scalar left, turned out to be a constant in this second version. It is important to notice that neither on the right hand side of (\ref{FSE}) or (\ref{FSE2}) we are expanding in terms of unit vectors. Therefore the names first, second and third normals or curvatures do not apply strictly.

\subsection{Fermi-Walker projections}
\label{FWP}

Let us focus now on equation (\ref{FSESIMPALPHA}). We can rewrite this equation as,

\begin{eqnarray}
u^{\alpha}_{\:\:\:;\mu}\:( A\: U^{\mu}_{(\phi_{1})} + B\: W^{\mu}_{(\varphi_{1})}) &=& {\delta u^{\alpha} \over \delta s} \nonumber\\  &=& -(q/m)\:\sqrt{-Q/2}\:\cos(\alpha) \:u_{\beta}\: \left(U^{\alpha}\:V^{\beta}- U^{\beta}\:V^{\alpha}\right) + \nonumber\\
&&(q/m)\:\sqrt{-Q/2}\:\sin(\alpha)\:u_{\beta}\: \left(Z^{\alpha}\:W^{\beta}- Z^{\beta}\:W^{\alpha}\right) \label{FWDECOMP} \ .
\end{eqnarray}

The two vectors $U^{\mu}_{(\phi_{1})}$ and $W^{\mu}_{(\varphi_{1})}$ represent a boost and a rotation as in equations (\ref{UT}) and (\ref{WT}) but in this case we are assuming that the angles $\phi_{1}$ and $\varphi_{1}$ define at every point along the curve the unit tangents on blades one and two respectively of the projections on both blades of the unit vector $u^{\alpha}$. That is to say $u^{\alpha} = A\: U^{\mu}_{(\phi_{1})} + B\: W^{\mu}_{(\varphi_{1})}$ such that the two local scalars $A = -u_{\alpha}\:U^{\alpha}_{(\phi_{1})}$ and $B = u_{\alpha}\:W^{\alpha}_{(\phi_{1})}$ satisfy $-A^{2} + B^{2} = -1$. Since we already know by gauge invariance that
\begin{eqnarray}
U^{[\alpha}_{(\phi_{1})}\:V^{\beta]}_{(\phi_{1})} = U^{[\alpha}\:V^{\beta]} = U^{[\alpha}_{(\phi)}\:V^{\beta]}_{(\phi)}  \label{Gaugeinv1}\\
Z^{[\alpha}_{(\varphi_{1})}\:W^{\beta]}_{(\varphi_{1})} = Z^{[\alpha}\:W^{\beta]} = Z^{[\alpha}_{(\varphi)}\:W^{\beta]}_{(\varphi)}  \label{Gaugeinv2} \ ,
\end{eqnarray}

we are able to rewrite equation (\ref{FWDECOMP}) as,

\begin{eqnarray}
u^{\alpha}_{\:\:\:;\mu}\:( A\: U^{\mu}_{(\phi_{1})} + B\: W^{\mu}_{(\varphi_{1})}) &=& {\delta u^{\alpha} \over \delta s} \nonumber\\  &=& -(q/m)\:\sqrt{-Q/2}\:\cos(\alpha) \:u_{\beta}\: \left(U^{\alpha}_{(\phi_{1})}\:V^{\beta}_{(\phi_{1})}- U^{\beta}_{(\phi_{1})}\:V^{\alpha}_{(\phi_{1})}\right) + \nonumber\\
&&(q/m)\:\sqrt{-Q/2}\:\sin(\alpha)\:u_{\beta}\: \left(Z^{\alpha}_{(\varphi_{1})}\:W^{\beta}_{(\varphi_{1})}- Z^{\beta}_{(\varphi_{1})}\:W^{\alpha}_{(\varphi_{1})}\right) \label{FWDECOMP2} \ .
\end{eqnarray}

If we take a look at equations (\ref{FWT}-\ref{FWS}), it is clear that we might interpret the right hand side of equation (\ref{FWDECOMP2}) as Fermi-Walker transport of $u^{\alpha}$ along a timelike curve with tangent $U^{\alpha}_{(\phi_{1})}$, and $b = -(q/m)\:\sqrt{-Q/2}\:\cos(\alpha)$ on one hand, and Fermi-Walker transport of $u^{\alpha}$ along a spacelike curve with tangent $W^{\alpha}_{(\varphi_{1})}$, and $d = -(q/m)\:\sqrt{-Q/2}\:\sin(\alpha)$ on the other hand. The Fermi-Walker projections on blades one and two. The right hand side of the general expression (\ref{FWDECOMP2}) is dictated by the superposition of Fermi-Walker transport on blades one and two. It represents a generalized Fermi-Walker equation for the problem of geometrodynamics. Through gauge invariance arguments we found the Lorentz force equation as a ``generalized'' Fermi-Walker transport.

\section{A particular case}
\label{anotherview}

We are going to consider a curve in spacetime, such that at each point, the projection of its tangent vector on blade one is given by ${dx_{1}^{\alpha}(s) \over ds}$, and on blade two by ${dx_{2}^{\alpha}(s) \over ds} = 0$. It is our objective to find special congruences of solutions to the Lorentz force equation, if they exist at all. Then, we are going to proceed as follows. For a ``generic'' scalar ``angle'' $\phi_{1}$ on blade one, and a ``generic'' scalar ``angle'' $\varphi_{1}$ on blade two, we are going to study the Frenet-Serret formulae associated to the timelike curve on blade one, whose tangent at every point is given by the unit timelike vector $U^{\alpha}_{(\phi_{1})} = {dx_{1}^{\alpha}(s) \over ds}= (u\:U^{\alpha}_{(\phi)} + v\:V^{\alpha}_{(\phi)})/(\:\sqrt{u^2-v^2})$ keeping the different notations consistent. Therefore, we are going to analyze the Frenet-Serret formulae for the tetrad $(U^{\alpha} _{(\phi_{1})},\:V^{\alpha} _{(\phi_{1})},\:Z^{\alpha} _{(\varphi_{1})},\:W^{\alpha} _{(\varphi_{1})})$. For the sake of simplicity we are going to assume that the transformation of the two vectors $(U^{\alpha},\:V^{\alpha})$ on blade one, given in (\ref{UTAV}-\ref{VTAV}) by the ``angle'' $\phi_{1}$ is a proper transformation. For improper transformations the result follows the same lines \cite{A}. Therefore we can write,

\begin{eqnarray}
U^{\alpha}_{(\phi_{1})}  &=& \cosh(\phi_{1})\: U^{\alpha} +  \sinh(\phi_{1})\: V^{\alpha} \label{UTAV} \\
V^{\alpha}_{(\phi_{1})} &=& \sinh(\phi_{1})\: U^{\alpha} +  \cosh(\phi_{1})\: V^{\alpha} \label{VTAV} \ .
\end{eqnarray}

The transformation of the two tetrad vectors $(Z^{\alpha},\:W^{\alpha})$ on blade two, given in (\ref{ZTAV}-\ref{WTAV}) by the ``angle'' $\varphi_{1}$, can be expressed as,

\begin{eqnarray}
Z^{\alpha}_{(\varphi_{1})}  &=& \cos(\varphi_{1})\: Z^{\alpha} -\sin(\varphi_{1})\: W^{\alpha} \label{ZTAV} \\
W^{\alpha}_{(\varphi_{1})}  &=& \sin(\varphi_{1})\: Z^{\alpha} +  \cos(\varphi_{1})\: W^{\alpha} \label{WTAV} \ .
\end{eqnarray}

We proceed then, to write the Frenet-Serret expressions associated to the timelike curve on blade one $x_{1}^{\alpha}(s)$ whose tangent is $U^{\alpha}_{(\phi_{1})}$,

\begin{eqnarray}
{\delta U^{\alpha}_{(\phi_{1})} \over \delta s}  &=& a_{1} \: V^{\alpha}_{(\phi_{1})} + b_{1}\: Z^{\alpha}_{(\varphi_{1})} + q_{1}\:W^{\alpha}_{(\varphi_{1})}   \label{FSU1}\\
{\delta V^{\alpha}_{(\phi_{1})}  \over \delta s} &=& a_{1} \: U^{\alpha}_{(\phi_{1})}  + p_{1}\: Z^{\alpha}_{(\varphi_{1})} + d_{1}\: W^{\alpha}_{(\varphi_{1})} \label{FSV1}\\
{\delta Z^{\alpha}_{(\varphi_{1})}  \over \delta s} &=& c_{1} \: W^{\alpha}_{(\varphi_{1})}  - p_{1} \: V^{\alpha}_{(\phi_{1})} +  b_{1} \: U^{\alpha}_{(\phi_{1})}   \label{FSZ1}\\
{\delta W^{\alpha}_{(\varphi_{1})}  \over \delta s} &=& -c_{1} \: Z^{\alpha}_{(\varphi_{1})} - d_{1} \: V^{\alpha}_{(\phi_{1})} + q_{1}\:U^{\alpha}_{(\phi_{1})}  \label{FSW1} \ .
\end{eqnarray}

The functions $a_{1}$, $b_{1}$, $c_{1}$, $d_{1}$, $p_{1}$ and $q_{1}$ are local scalars. In addition, we are also going to introduce the auxiliary scalar variables,
\begin{eqnarray}
A_{z} &=& Z_{\alpha}\:U^{\alpha}_{\:\:\:;\beta}\:U^{\beta} \label{ASz} \\
B_{z} &=& Z_{\alpha}\:U^{\alpha}_{\:\:\:;\beta}\:V^{\beta} \label{BSz} \\
C_{z} &=& Z_{\alpha}\:V^{\alpha}_{\:\:\:;\beta}\:U^{\beta} \label{CSz} \\
D_{z} &=& Z_{\alpha}\:V^{\alpha}_{\:\:\:;\beta}\:V^{\beta} \label{DSz} \\
A_{w} &=& W_{\alpha}\:U^{\alpha}_{\:\:\:;\beta}\:U^{\beta} \label{ASw} \\
B_{w} &=& W_{\alpha}\:U^{\alpha}_{\:\:\:;\beta}\:V^{\beta} \label{BSw} \\
C_{w} &=& W_{\alpha}\:V^{\alpha}_{\:\:\:;\beta}\:U^{\beta} \label{CSw} \\
D_{w} &=& W_{\alpha}\:V^{\alpha}_{\:\:\:;\beta}\:V^{\beta} \label{DSw} \ .
\end{eqnarray}

Now comes the core on the idea to find this special solution to the Lorentz force equation. Let us demand or impose in (\ref{FSU1}) the following equations,

\begin{eqnarray}
Z_{\alpha(\varphi_{1})} \:{\delta U^{\alpha}_{(\phi_{1})} \over \delta s} &=& 0 \label{eq1b1} \\
W_{\alpha(\varphi_{1})} \:{\delta U^{\alpha}_{(\phi_{1})} \over \delta s} &=& 0 \ .\label{eq2b1}
\end{eqnarray}

Let us remember that we are assuming that the curve has null projection on blade two. These equations (\ref{eq1b1}-\ref{eq2b1}) represent equations for the two local scalar functions $(\phi_{1},\varphi_{1})$. If we replace in (\ref{eq1b1}-\ref{eq2b1}) expressions (\ref{UTAV}-\ref{VTAV}) and (\ref{ZTAV}-\ref{WTAV}), work out the algebraic equations, and make use of (\ref{ASz}-\ref{DSw}) we find,

\begin{eqnarray}
\tan(\varphi_{1}) &=& {\left(\cosh^{2}(\phi_{1})\:A_{z} + \sinh^{2}(\phi_{1})\:D_{z} + \cosh(\phi_{1})\:\sinh(\phi_{1})\:(B_{z} + C_{z})\right) \over \left(\cosh^{2}(\phi_{1})\:A_{w} + \sinh^{2}(\phi_{1})\:D_{w} + \cosh(\phi_{1})\:\sinh(\phi_{1})\:(B_{w} + C_{w})\right)} \label{eqvarphi1b1}
\end{eqnarray}
\begin{eqnarray}
\tan(\varphi_{1}) &=& - { \left(\cosh^{2}(\phi_{1})\:A_{w} + \sinh^{2}(\phi_{1})\:D_{w} + \cosh(\phi_{1})\:\sinh(\phi_{1})\:(B_{w} + C_{w})\right) \over \left(\cosh^{2}(\phi_{1})\:A_{z} + \sinh^{2}(\phi_{1})\:D_{z} + \cosh(\phi_{1})\:\sinh(\phi_{1})\:(B_{z} + C_{z})\right) } \ .\label{eqvarphi2b1}
\end{eqnarray}

Let us remember that the absolute derivatives are taken relative to vector (\ref{UTAV}). Now, in order for both equations (\ref{eqvarphi1b1}-\ref{eqvarphi2b1}) to be compatible, it can be readily noticed that the following equality would have to be true, $\tan(\varphi_{1}) = - 1 / \tan(\varphi_{1})$. Then, there are no solutions to the ``pure'' blade one Lorentz force equation problem. We must also remind ourselves that in this paper we are dealing with non-null electromagnetic fields, therefore the local scalar $\tan(2\alpha)$ is not null or zero \cite{A}$^{,}$\cite{MW}, which means that $f_{\mu\nu}$ has a non-null projection on both blades, see equation (\ref{EMF}).

\section{A more general example}
\label{anothergenview}

It is our objective to find special congruences of solutions to the Lorentz force equation, as a more general example using our geometrical techniques. We are increasing the level of complexity following a pedagogical scheme. Thus, we are going to proceed as follows. For a ``generic'' scalar ``angle'' $\phi_{1}$ on blade one, and a ``generic'' scalar ``angle'' $\varphi_{1}$ on blade two, we are going to study the Frenet-Serret formulae associated to the timelike curve, whose projection on blade one at every point is given by the timelike vector $(M / \sqrt{M^2 - N^2}\:)\:U^{\alpha}_{(\phi_{1})} = (M / \sqrt{M^2 - N^2}\:)\: (u\:U^{\alpha}_{(\phi)} + v\:V^{\alpha}_{(\phi)})/\sqrt{u^2-v^2} $ and its projection on blade two is given by $(N / \sqrt{M^2 - N^2}\:)\:W^{\alpha}_{(\varphi_{1})} = (N / \sqrt{M^2 - N^2}\:)\: (z\:Z^{\alpha}_{(\varphi)} + w\:W^{\alpha}_{(\varphi)})/\sqrt{z^2+w^2} $. Both $M$ and $N$ are assumed to be non-trivial local scalars. We have already analyzed the Frenet-Serret formulae for the tetrad $(U^{\alpha} _{(\phi_{1})},\:V^{\alpha} _{(\phi_{1})},\:Z^{\alpha} _{(\varphi_{1})},\:W^{\alpha} _{(\varphi_{1})})$. We are simply going to take advantage of this previous analysis to find solutions to the Lorentz force equation of a more general nature. For brevity we are going to call $A = M / \sqrt{M^2 - N^2}$ and $B = N / \sqrt{M^2 - N^2}$. We proceed then, to write the Frenet-Serret expressions associated to the projections of the curve both on blade one and two,

\begin{eqnarray}
{\delta \left( A\:U^{\alpha}_{(\phi_{1})}\right) \over \delta s}  &=& {\delta \left( A \right) \over \delta s}\: U^{\alpha}_{(\phi_{1})} + A\:{\delta U^{\alpha}_{(\phi_{1})} \over \delta s}\label{GEN1}\\
{\delta U^{\alpha}_{(\phi_{1})} \over \delta s}  &=& a_{1} \: V^{\alpha}_{(\phi_{1})} + b_{1}\: Z^{\alpha}_{(\varphi_{1})} + q_{1}\:W^{\alpha}_{(\varphi_{1})}   \label{GEN11}\\
{\delta V^{\alpha}_{(\phi_{1})}  \over \delta s} &=& a_{1} \: U^{\alpha}_{(\phi_{1})}  + p_{1}\: Z^{\alpha}_{(\varphi_{1})} + d_{1}\: W^{\alpha}_{(\varphi_{1})} \label{GEN2}\\
{\delta Z^{\alpha}_{(\varphi_{1})}  \over \delta s} &=& c_{1} \: W^{\alpha}_{(\varphi_{1})}  - p_{1} \: V^{\alpha}_{(\phi_{1})} +  b_{1} \: U^{\alpha}_{(\phi_{1})}   \label{GEN3}\\
{\delta W^{\alpha}_{(\varphi_{1})}  \over \delta s} &=& -c_{1} \: Z^{\alpha}_{(\varphi_{1})} - d_{1} \: V^{\alpha}_{(\phi_{1})} + q_{1}\:U^{\alpha}_{(\phi_{1})}  \label{GEN44}\\
{\delta \left( B\:W^{\alpha}_{(\varphi_{1})}\right)  \over \delta s} &=& {\delta \left( B \right)  \over \delta s}\: W^{\alpha}_{(\varphi_{1})} + B\:{\delta W^{\alpha}_{(\varphi_{1})}  \over \delta s} \label{GEN4} \ .
\end{eqnarray}

The functions $a_{1}$, $b_{1}$, $c_{1}$, $d_{1}$, $p_{1}$ and $q_{1}$ are again local scalars.

Once more comes the core on the idea to find these more general solutions to the Lorentz force equation. Let us demand or impose in (\ref{GEN1}-\ref{GEN4}) the following equations,

\begin{eqnarray}
{\delta \left( A \right) \over \delta s} &=& 0 \label{SCALAR1} \\
{\delta \left( B \right)  \over \delta s} &=& 0 \label{SCALAR2} \\
W_{\alpha(\varphi_{1})} \:{\delta U^{\alpha}_{(\phi_{1})} \over \delta s} &=& 0 \ .\label{ORTHO2}
\end{eqnarray}

We know of course that either of (\ref{SCALAR1}) or (\ref{SCALAR2}) would be enough since $-A^{2}+B^{2}=-1$.
The problem is posed as follows. First, we are supposed to know the tetrad vectors and their covariant derivatives from solutions to the Einstein-Maxwell equations (\ref{EM1}-\ref{EM3}). Second, we can supplement equations (\ref{SCALAR1}-\ref{ORTHO2}) with another one in order to find $(M, N)$ and $(\phi_{1},\varphi_{1})$. For instance, $Z_{\alpha(\varphi_{1})} \:{\delta V^{\alpha}_{(\phi_{1})} \over \delta s} = 0$. Each possible choice for an additional equation yields different solutions $(M, N)$ and $(\phi_{1},\varphi_{1})$ to the Lorentz force equation. We finally see that Frenet-Serret equations (\ref{GEN11}) and (\ref{GEN44}) after imposing the supplementary equations turn out to be,

\begin{eqnarray}
{\delta U^{\alpha}_{(\phi_{1})} \over \delta s}  &=& a_{1} \: V^{\alpha}_{(\phi_{1})} + b_{1}\: Z^{\alpha}_{(\varphi_{1})} \label{RED11} \\
{\delta W^{\alpha}_{(\varphi_{1})}  \over \delta s} &=& -c_{1} \: Z^{\alpha}_{(\varphi_{1})} - d_{1} \: V^{\alpha}_{(\phi_{1})}   \label{RED44}
 \ .
\end{eqnarray}

Our intention is to study the absolute derivative of the timelike vector $A\:U^{\alpha}_{(\phi_{1})} + B \: W^{\alpha}_{(\varphi_{1})}$ along a timelike integral curve whose tangent is precisely $A\:U^{\alpha}_{(\phi_{1})} + B \: W^{\alpha}_{(\varphi_{1})}$,

\begin{eqnarray}
{\delta \left( A\: U^{\alpha}_{(\phi_{1})} + B\: W^{\alpha}_{(\varphi_{1})} \right) \over \delta s} &=& A \: {\delta U^{\alpha}_{(\phi_{1})} \over \delta s} + B \: {\delta W^{\alpha}_{(\varphi_{1})}  \over \delta s}   \ , \label{GENEQ1}
\end{eqnarray}

because of equations (\ref{SCALAR1}-\ref{SCALAR2}). Now, using equations (\ref{RED11}-\ref{RED44}) we can write,

\begin{eqnarray}
{\delta \left( A\: U^{\alpha}_{(\phi_{1})} + B\: W^{\alpha}_{(\varphi_{1})} \right) \over \delta s} &=& A \: \left( a_{1} \: V^{\alpha}_{(\phi_{1})} + b_{1}\: Z^{\alpha}_{(\varphi_{1})} \right) + B \: \left( -c_{1} \: Z^{\alpha}_{(\varphi_{1})} - d_{1} \: V^{\alpha}_{(\phi_{1})} \right)  \ . \label{GENEQ2}
\end{eqnarray}

It is simple to see that equation (\ref{GENEQ2}) can be compared to, or rewritten as,

\begin{eqnarray}
\lefteqn{{\delta \left( A\: U^{\alpha}_{(\phi_{1})} + B\: W^{\alpha}_{(\varphi_{1})} \right) \over \delta s} = (q/m) \: \left( A \: U_{\rho(\phi_{1})} + B\: W_{\rho(\varphi_{1})} \right)} \nonumber \\
&&\sqrt{-Q/2}\:\:\left( - \cos\alpha\:\: \left( U^{\alpha}_{(\phi_{1})}\:V^{\rho}_{(\phi_{1})} - U^{\rho}_{(\phi_{1})}\:V^{\alpha}_{(\phi_{1})} \right) + \sin\alpha\:\:\left( Z^{\alpha}_{(\varphi_{1})}
\:W^{\rho}_{(\varphi_{1})} - Z^{\rho}_{(\varphi_{1})}
\:W^{\alpha}_{(\varphi_{1})} \right)\right). \label{GENEQ3}
\end{eqnarray}

as long as we ask for the scalar functions $a_{1}$, $b_{1}$, $c_{1}$ and $d_{1}$ to satisfy the following algebraic equations,

\begin{eqnarray}
-\:\sqrt{-Q/2}\:\:\cos\alpha\:A\:q/m &=& A\:a_{1} - B\:d_{1} \label{algeb1}\\
\:\sqrt{-Q/2}\:\:\sin\alpha\:B\:q/m &=& A\:b_{1} - B\:c_{1} \label{algeb2}\ .
\end{eqnarray}

We can also remind ourselves of the gauge invariant equalities \cite{A},

\begin{eqnarray}
U^{[\alpha}_{(\phi_{1})}\:V^{\beta]}_{(\phi_{1})} = U^{[\alpha}\:V^{\beta]} = U^{[\alpha}_{(\phi)}\:V^{\beta]}_{(\phi)}  \label{GEQphi1}\\
Z^{[\alpha}_{(\varphi_{1})}\:W^{\beta]}_{(\varphi_{1})} = Z^{[\alpha}\:W^{\beta]} = Z^{[\alpha}_{(\varphi)}\:W^{\beta]}_{(\varphi)}  \label{GEQvarphi2} \ .
\end{eqnarray}

Therefore, equation (\ref{GENEQ3}) can be expressed as,

\begin{eqnarray}
\lefteqn{{\delta \left( A\: U^{\alpha}_{(\phi_{1})} + B\: W^{\alpha}_{(\varphi_{1})} \right) \over \delta s} =}\nonumber\\
&&(q/m)\:\left( A \: U_{\rho(\phi_{1})} + B\: W_{\rho(\varphi_{1})} \right)
\left( -2\:\sqrt{-Q/2}\:\:\cos\alpha\:\:U^{[\alpha}\:V^{\rho]} + 2\:\sqrt{-Q/2}\:\:\sin\alpha\:\:Z^{[\alpha}\:W^{\rho]}\right)\:\:\: \label{GENEQ4}
\end{eqnarray}

Calling $u^{\alpha} = A\: U^{\alpha}_{(\phi_{1})} + B\: W^{\alpha}_{(\varphi_{1})}$, and reminding equation (\ref{EMF}), we can rewrite (\ref{GENEQ4}) as,

\begin{equation}
{\delta u^{\alpha} \over \delta s} = (q/m)\: u_{\rho}\: f^{\alpha\rho}\ , \label{Lorentz}\\
\end{equation}

that is, the Lorentz ``force'' equation for the particle of mass $m$ and charge $q$ for the timelike vector $u^{\alpha} = A\: U^{\alpha}_{(\phi_{1})} + B\: W^{\alpha}_{(\varphi_{1})}$.

\section{Conclusions}

The introduction of new tetrads for non-null electromagnetic fields in \cite{A} allows for the geometrical analysis of the Lorentz force equation. Through Frenet-Serret expressions it is possible to identify curvature scalars in terms of the local scalar functions $\alpha$ and $\sqrt{-Q/2}$ associated to the non-null electromagnetic field present in a curved spacetime. It is a fundamental result we understand, to deduce the Lorentz force equation on purely geometrical and gauge invariance arguments as a generalized form of Fermi-Walker transport. It was never done before. We proved that the Lorentz force equation can be deduced through Riemannian geometry arguments. Only scalars had to be chosen. And in the second version \ref{FSsv} just a constant. The ratio $q/m$ is proved to be nothing but a curvature scalar (not the first curvature strictly speaking as was stated above) asociated to the timelike trajectory of a charged particle. But a geometrical scalar, in the Riemannian sense. Deep in the theory and arising from the results in \cite{A} is the fact that the stress-energy tensor in geometrodynamics (\ref{EM3}) is a geometrical object that by itself defines two blades at every point in spacetime and the vectors that diagonalize this tensor can be found explicitly through the use of the extremal field and electromagnetic gauge vectors. A blade one ``pure'' solutions are studied and found not to exist through the use of Frenet-Serret expressions. Standard formulations in electromagnetism never even addressed this issue. It is not even possible to be aware of this situation using standard tools. More general solutions are found with non-trivial projections on both blades. These are new techniques available now to analyze the solutions to the Lorentz force equation from a geometrical point of view. It is like geometrical ``reverse engineering''. The new tetrads allow for a better insight regarding the relationship between traditional Abelian gauge theories on one hand, and traditional Riemannian geometry on the other hand. The key lies upon the two theorems proved in \cite{A} relating through isomorphisms the groups $U(1)$ and LB1, LB2. The local ``internal'' transformations that leave invariant the electromagnetic field are proved to be isomorphic to the local tetrad transformations either on blade one or two. In other words, the local ``internal'' group of transformations is isomorphic to a subgroup of the local group of Lorentz transformations. This was not known before. It is precisely this ``gauge geometry'' the tool that allows for a new understanding of a traditional equation like the Lorentz force equation. We rewrote the usual theory of electromagnetic fields in curved spacetime in a new language. The new tetrads replace the standard variables. The gauge transformations of the potentials are reinterpreted through local group isomorphisms as tetrad Lorentz transformations on both blades. In fact this is the way to reexpress the standard gauge theories into a geometrical Riemannian language. The goal is to show that there is a whole new geometrical way to understand or interpret the gauge theories, the equations involved and their solutions under a whole new light. Other future applications are possible. For instance, the study of the kinematics in these spacetimes \cite{RG}$^{,}$\cite{RW}$^{,}$\cite{MC2}$^{,}$\cite{EJKS}, that is, the search for a ``connection'' to the theory of embeddings, time slicings \cite{KK}$^{,}$\cite{JDBKK}$^{,}$\cite{FM}$^{,}$\cite{FAEP}, the initial value formulation \cite{JWY1}$^{,}$\cite{JWY2}$^{,}$\cite{NOMJWY}$^{,}$\cite{HY}$^{,}$\cite{JY}$^{,}$\cite{LICH}$^{,}$\cite{YCB}$^{,}$\cite{CMDWYCB}, the Cauchy evolution \cite{ADM}$^{,}$\cite{EB}$^{,}$\cite{ERW}$^{,}$\cite{LSJWY}, etc. Simplicity is the most sought after quality in all these geometrical descriptions, and the previously introduced tetrads, blades, isomorphisms, prove to be practical and useful. We quote from \cite{MW} ``The question poses itself insistently to find a point of view which will make Rainich-Riemannian geometry seem a particularly natural kind of geometry to consider''.

\acknowledgements

This work was partially funded by PEDECIBA.

\end{document}